\documentstyle[11pt]{article}
\def\Journal#1#2#3#4{{#1} {\bf #2}, #3 (#4)}

\def\NIMA{{\em Nucl. Instrum. Methods} A}
\def\NPB{{\em Nucl. Phys.} B}
\def\PLB{{\em Phys. Lett.}  B}
\def\PRL{\em Phys. Rev. Lett.}
\def\PRD{{\em Phys. Rev.} D}

\def\NPA{{\em Nucl. Phys.} A}
\def\PRC{{\em Phys. Rev.} C}
\def\PR{\em Phys. Rep.}
\begin{document}
\begin{titlepage}

\hspace{11cm}BIHEP-TH-99-10

\vspace{1cm}

\centerline{\large \bf POSSIBLE $\Delta\Delta$ DIBARYONS IN THE QUARK CLUSTER
MODEL
\footnote{This work was supported in part by the National
Natural Science Foundation of China under grant number 1977551B and the Chinese
Academy of Sciences under grant number B78}}
\vspace{1cm}
\centerline{ Q.B.Li$^{a}$, P.N.Shen$^{a,b,c,d}$}
\vspace{1cm}
{\small
{
\flushleft{\bf  $~~~$a. Institute of High Energy Physics, Chinese Academy
of Sciences, P.O.Box 918(4),}
\flushleft{\bf  $~~~~~~$Beijing 100039, China}
\vspace{8pt}
\flushleft{\bf  $~~~b.$ China Center of Advanced Science and Technology
 (World Laboratory),}
\flushleft{\bf  $~~~~~~$P.O.Box 8730, Beijing 100080, China}
\vspace{8pt}
\flushleft{\bf  $~~~c.$ Institute of Theoretical Physics, Chinese Academy
of Sciences, P.O.Box 2735,}
\flushleft{\bf  $~~~~~~$Beijing 100080, China}
\vspace{8pt}
\flushleft{\bf  $~~~d.$ Center of Theoretical Nuclear Physics, National Lab of
Heavy Ion Accelerator, }
\flushleft{\bf  $~~~~~~$Lanzhou 730000, China}
\vspace{8pt}
}}

\vspace{2cm}

\centerline{\bf Abstract}
In the framework of $RGM$, the binding energy of one channel
$\Delta\Delta_{(3,0)}$($d^*$) and
$\Delta\Delta_{(0,3)}$ are studied in the chiral $SU(3)$
quark cluster model. It is shown that the binding energies of the systems
are a few tens of MeV. The behavior of the chiral field is also investigated
by comparing the results with those in the $SU(2)$ and the extended
$SU(2)$ chiral quark models. It is found that the symmetry property of the $\Delta\Delta$ system makes the contribution of the relative kinetic energy 
operator between two clusters attractive. This is very beneficial for forming
the bound dibaryon. Meanwhile the chiral-quark field coupling also plays a very
important role on binding. The S-wave phase shifts and the corresponding
scattering lengths of the systems are also given.
 
\noindent

\end{titlepage}

\baselineskip 18pt

\section{Introduction}

\vspace{0.3cm}

With the consideration that the Quantum Chromodynamics (QCD) is 
the underlying theory of the strong interaction,
the hadronic systems are assumed to be composed of quarks and gluons, and 
most of descriptions of baryons are associated with some effective degrees of
freedom of nonperturbative $QCD$ (NPQCD), such as constituent quarks [1],
and chiral fields [2], etc.. But, in the multi-baryon system, the property
of the system mostly can be well explained in the baryon and meson degrees of
freedom with only some special cases where the quark and gluon degrees of freedom has to be 
taken into account as exceptions[3]. In principle, nothing special would prohibit a system
with the baryon number larger than one, and there might exist dibaryons 
which are composed of six quarks. Many efforts have been made in both theoretical and experimental investigations to search such object in past 
years [4-7]. Unlike the deuteron, the dibaryon has a QCD motivated
origin and should be a system with quarks and gluons confining in a 
rather smaller volume, say smaller than 0.85fm in radius.
In this system, both the perturbative QCD (PQCD) and the NPQCD effects
should be taken into account. However, because of the complexity of NPQCD,
one has to use effective models to simulate the strong interaction, 
especially in the NPQCD region. Thus, investigating dibaryons is
a prospective field to test various models of NPQCD, and consequently,
to enrich our knowledge of the QCD phenomenology.

\vspace{0.3cm}

According to Jaffe's calculation with the bag model [4], the color magnetic
interaction (CMI) of the one-gluon-exchange (OGE) potential of the H particle    
is attractive. In 1987, K.Yazaki [8] investigated non-strange six-quark 
systems by considering OGE and confinement potentials and showed that 
CMI in the $\Delta\Delta_{(3,0)}$ system demonstrate the attractive feature among the $NN$, $N\Delta$ and $\Delta\Delta$ systems. Therefore, this state 
Is highly possible a dibaryon. Studying this system is particularly plausible.

\vspace{0.3cm} 

Since Jaffe predicted the H dibaryon in 1977 [4], many works have been done
on searching the possible existences of dibaryons such as strangenessless 
dibaryons $d^*$ [5,8-10] and $d'$ [11] and the strange dibaryons such as  
H, $\Omega\Omega$, $\Xi^*\Omega$ and $\Xi\Omega$ [13,14,12].
 The current calculations of H showed that the 
mass of H is around the threshold of the $\Lambda\Lambda$ channel [14,15].
This result is consistent with the reports of current experiment [16].
The recent theoretical result of $d^*$ is obtained by considering the coupling 
of the $\Delta\Delta$ and $CC$(hidden color) channels in the chiral quark
models. It is shown that the binding energy of $d^*$ is in the order of tens 
MeV [10]. The predicted $\Omega\Omega$, $\Xi^{*}\Omega$ and $\Xi\Omega$ are
about 100 MeV, 80 MeV and 30 MeV, respectively[12, 23-25].

\vspace{0.3cm}

In the theoretical study of dibaryon, the quark cluster models have been 
Intensively employed. With these models, one can easily realize the
antisymmetrization character between quarks belonging to different clusters,
which can only be neglected when clusters are well separated from each other.

\vspace{0.3cm}

For the system with a radius not larger than 0.85fm, the Pauli
principle on the quark level would play a substantial role [20].
According to our calculation, $\Delta\Delta_{(3,0)}$ $(d^*)$ has the most
antisymmetized property in the spin-flavor-color space, which is characterized 
by the expectation value of the antisymmetrizer in the spin-flavor-color
space. This property appears in another non-strange six-quark system,
$\Delta\Delta_{(0,3)}$. 
In this paper, we would discuss the one channel $\Delta\Delta_{(0,3)}$
together with $d^*$ within the chiral quark cluster model.
The paper is organized in such a way that the model is briefly presented 
in Sect.2, the results and discussions are given in Sect.3 and the 
conclusions is drawn in Sect.4.

\noindent

\section{The effective Hamiltonian and two-cluster wavefunction}

\noindent

    As an effective theory of QCD, in the chiral quark model, the 
constituent quark and the chiral field are assumed as the effective degrees 
of freedom, and a long-range confinement term, which dominants the long-range 
NPQCD effect and still cannot strictly be derived up to now, a 
short-range one-gluon-exchange term, which describes the short-range QCD
effect and a set of terms induced by the coupling between the quark and 
chiral fields, which mainly depict the short- and medium-range NPQCD effects
are considered as the effective interaction. Thus the dynamics of a six-quark
system in the SU(3) chiral quark model is governed by the effective 
Hamiltonian [18,19]

\begin{eqnarray}
H~=~T~+~\sum_{i<j}\big(V_{ij}^{CONF}~+~V_{ij}^{OGE}~+~V_{ij}^{PS}~+~
V_{ij}^{S}\big),
\end{eqnarray}
where $T$ denotes the kinetic energy operator of the system and
$V_{ij}^{CONF}$, $V_{ij}^{OGE}$, $V_{ij}^{PS}$ and $V_{ij}^{S}$
represent the confinement, one-gluon exchange, pseudo-scalar chiral field
induced and scalar chiral field induced potentials intervening between the
$i-$th and $j-$th quarks, respectively. The confinement potential can 
phenomenologically take a quadratic form

\begin{eqnarray}
V_{ij}^{CONF}~=~-(\lambda^{a}_{i} \lambda^{a}_{j})_{c}(a_{ij}^{0}
+a_{ij}r_{ij}^{2}),
\end{eqnarray}
where the superscript a is the color index. The OGE potential can be 
derived from the perturbative tree diagram

\begin{eqnarray}
V_{ij}^{OGE}~=~\frac{g_{i} g_{j}}{4}(\lambda^{a}_{i} \lambda^{a}_{j})_{c}
~\lbrack~ \frac{1}{r_{ij}}&-&\frac{\pi}{2}\big(\frac{1}{m_{i}^{2}}+\frac{1}
{m_{j}^{2}}+\frac{4}{3}\frac{\vec{\sigma}_{i}\cdot\vec{\sigma}_{j}}
{m_{i}m_{j}}\big)\delta(\vec{r}_{ij}) \\ \nonumber
     &-&\frac{1}{4m_{i}m_{j}r_{ij}^{3}}S_{ij}-\frac{3}
     {4m_{i}m_{j}r_{ij}^{3}}\vec{L}\cdot(\vec{\sigma}_{i}+\vec{\sigma}_{j})
     ~\rbrack,
\end{eqnarray}
with the tensor operator $S_{ij}$ being
\begin{eqnarray}
S_{ij}~&=&~3(\vec{\sigma}_{i}\cdot\hat{r}_{ij})~
(\vec{\sigma}_{j}\cdot\hat{r}_{ij}) - (\vec{\sigma}_{i}
\cdot\vec{\sigma}_{j})~.
\end{eqnarray}
The pseudo-scalar and scalar field induced potentials are originated from the restoration of the important symmetry of strong interaction---the
chiral symmetry and can be written as 

\begin{eqnarray}
V_{ij}^{PS} &=& C(g_{ch},~m_{\pi_{a}},
~\Lambda)~\frac{m_{\pi_{a}}^{2}}{12m_{i}m_{j}}  \nonumber  \\
 &\cdot&\lbrack f_{1}(m_{\pi_{a}},~\Lambda,~r_{ij})~(\vec{\sigma_{i}}\cdot
 \vec{\sigma_{j}}) + f_{2}(m_{\pi_{a}},~\Lambda,~r_{ij})~S_{ij}~\rbrack
(\lambda^{a}_{i}\lambda^{a}_{j})_{f}~,
\end{eqnarray}
and 

\begin{eqnarray}
V_{ij}^{S} &=& - C(g_{ch},~m_{\sigma_{a}},
~\Lambda)  
 \cdot \lbrack f_{3}(m_{\sigma_{a}},~\Lambda,~r_{ij}) \nonumber \\
 &+& \frac{m_{\sigma_{a}}^{2}}{4m_{i}m_{j}}~
 f_{4}(m_{\sigma_{a}},~\Lambda,~r_{ij})~\vec{L}\cdot(\vec{\sigma_{i}}
 +\vec{\sigma_{j}})~\rbrack (\lambda^{a}_{i}\lambda^{a}_{j})_{f}~,
\end{eqnarray}
respectively. In above equations, the subscript $f$ denotes the flavor index.
The functions $f_{i},~Y,~G$, and $H$ and the constant c-number $C$ are

\begin{eqnarray}
f_{1}(m,~\Lambda,~r)~&=&~Y(mr) -
  \Big(\frac{\Lambda}{m}\Big)^{3}~Y(\Lambda r),\\
f_{2}(m,~\Lambda,~r)~&=&~H(mr) -
  \Big(\frac{\Lambda}{m}\Big)^{3}~H(\Lambda r),\\
f_{3}(m,~\Lambda,~r)~&=&~Y(mr) - \frac{\Lambda}{m}~Y(\Lambda r),\\
f_{4}(m,~\Lambda,~r)~&=&~G(mr) - \Big(\frac{\Lambda}{m}\Big)^{3}~G(\Lambda r),\\
C(g,~m,~\Lambda)~&=&~\frac{g_{ch}^{2}}{4\pi}~\frac{\Lambda^{2}~m}{\Lambda^{2}
-m^{2}},\\
\end{eqnarray}

respectively, with

\begin{eqnarray}
Y(mr)~&=&~\frac{1}{mr}~e^{-mr},\\
G(mr)~&=&~\frac{1}{mr}~\Big(1+\frac{1}{mr}\Big)~Y(mr),\\
H(mr)~&=&~~\Big(1+\frac{3}{mr}+\frac{3}{m^{2}r^{2}}\Big)~Y(mr),\\
\frac{g_{ch}^{2}}{4\pi}~&=&~\frac{9}{25}~\frac{g_{NN\pi}^{2}}{4\pi}~
\Big(\frac{m_{q}}{M_{N}}\Big)^{2}.
\end{eqnarray}

In Eq.(5), $\pi_{a}$ with ($a=1,2,...,8,$ and $0$) correspond
to the pseudoscalar fields $\pi,~K,~\eta_{8}$ and $\eta_{0}$, respectively,
and $\eta$ and $\eta'$ are the linear combinations of $\eta_{0}$ and
$\eta_{8}$ with the mixing angle $\theta$ [14]. In Eq.(6),
$\sigma_{a}$ with ($a=1,2,...,8,$ and $0$) denotes the  scalar
fields $\sigma',~\kappa,~\epsilon$ and $~\sigma$, respectively.

\vspace{0.3cm}

In the framework of Resonating Group Method $(RGM)$, the wavefunction of the
Six-quark system can be written as

\begin{eqnarray}
  {\Psi}_{6q}~=~{\cal A}[{\Phi}_A{\Phi}_B\chi({\bf R}_{AB})Z({\bf R}_{cm})]
\end{eqnarray}

where $\phi_{A(B)}$ denotes the wavefunction of cluster A(B), 
${\chi}({\bf R}_{AB})$ is the trial wavefunction between clusters A
and B, $Z({\bf R}_{cm})$ is the wavefunction of the center of mass motion (CM)
of the six quark system and $\cal A$ represents the antisymmetrizer. 
Expanding the unknown wavefunction ${\chi}({\bf R}_{AB})$ by well-defined 
basis functions, such as Gaussian functions, one can solve $RGM$ bound state
equation to obtain eigenvalues and corresponding wavefunctions, simultaneously. The details of solving $RGM$ bound-state problem can be found in Refs. [21,22].

\vspace{0.3cm}

The model parameters should be fixed at the very beginning by the mass
splittings among $N$,$\Delta$, $\Sigma$ and $\Xi$, respectively, and the 
stability conditions of octet $(S=1/2)$ and decuplet $(S=3/2)$ baryons,
respectively. The resultant values are listed in Table 1.

\centerline{\bf {Table 1~~~~Model parameters under $SU(3)$ chiral quark model}}
{\small
\begin{center}
\begin{tabular}{ccc|ccc}\hline
                              & $~~~~~$    Set1  & $~~~~~$  Set2 &                                   & $~~~~~$  Set1  & $~~~~~$  Set2    \\\hline
  $m_u~(MeV)$                 & $~~~~~$    313   & $~~~~~$  313  &                                   &                &                  \\
  $m_s~(MeV)$                 & $~~~~~$    470   & $~~~~~$  470  &                                   &                &                  \\
  $b_u~(fm)$                  & $~~~~~$    0.505 & $~~~~~$  0.505 &                                   &                &                  \\
  $m_\pi~(fm^{-1})$           & $~~~~~$    0.7   & $~~~~~$  0.7  & $\Lambda_\pi~(fm^{-1})$           & $~~~~~$  4.2   & $~~~~~$  4.2     \\
  $m_k~(fm^{-1})$             & $~~~~~$    2.51  & $~~~~~$  2.51 & $\Lambda_k~(fm^{-1})$             & $~~~~~$  4.2   & $~~~~~$  4.2     \\
  $m_\eta~(fm^{-1})$          & $~~~~~$    2.78  & $~~~~~$  2.78 & $\Lambda_\eta~(fm^{-1})$          & $~~~~~$  5.0   & $~~~~~$  5.0     \\
  $m_{\eta'}~(fm^{-1})$       & $~~~~~$    4.85  & $~~~~~$  4.85 & $\Lambda_{\eta'}~(fm^{-1})$       & $~~~~~$  5.0   & $~~~~~$  5.0     \\
  $m_\sigma~(fm^{-1})$        & $~~~~~$    3.17  & $~~~~~$  3.17 & $\Lambda_\sigma~(fm^{-1})$        & $~~~~~$  4.2   & $~~~~~$  7.0     \\
  $m_{\sigma'}~(fm^{-1})$     & $~~~~~$    4.85  & $~~~~~$  4.85 & $\Lambda_{\sigma'}~(fm^{-1})$     & $~~~~~$  5.0   & $~~~~~$  5.0     \\
  $m_\kappa~(fm^{-1})$        & $~~~~~$    4.85  & $~~~~~$  7.09 & $\Lambda_\kappa~(fm^{-1})$         & $~~~~~$  5.0   & $~~~~~$  7.61     \\
  $m_\epsilon~(fm^{-1})$      & $~~~~~$    4.85  & $~~~~~$  7.09 & $\Lambda_\epsilon~(fm^{-1})$      & $~~~~~$  5.0   & $~~~~~$  7.61     \\
  $g_u$                       & $~~~~~$  0.936   & $~~~~~$ 0.936 &                                   &                &                  \\
  $g_s$                       & $~~~~~$  0.924   & $~~~~~$ 0.781 &                                   &                &                  \\
  $a_{uu}~(MeV/fm^2)$         & $~~~~~$  54.34   & $~~~~~$ 57.71 & $a^0_{uu}~(MeV)$                  & $~~~~~$ -47.69 & $~~~~~$ -48.89   \\
  $a_{us}~(MeV/fm^2)$         & $~~~~~$  65.75   & $~~~~~$ 66.51 & $a^0_{us}~(MeV)$                  & $~~~~~$ -41.73 & $~~~~~$ -50.57   \\
  $a_{ss}~(MeV/fm^2)$         & $~~~~~$  102.97  & $~~~~~$ 115.39 & $a^0_{ss}~(MeV)$                  & $~~~~~$ -45.04 & $~~~~~$ -68.11   \\\hline               &

\end{tabular}
\end{center}
}
In this table, $m_A$ denotes the mass of particle A which can be
either valence quark or meson involved, $\Lambda_A$ represents
the corresponding cut-off mass of the particle, $g_q$ is the coupling constant
of gluon to the valence quark q, $a_{q_{1}q_{2}}$ depicts the confinement
strength between valence quarks $q_{1}$ and $q_{2}$, respectively,  and
$a^0_{q_{1}q_{2}}$ denotes the corresponding zero-point energy.
It should be mentioned again that with both sets of parameters, one can 
reasonably reproduce the NN and NY scattering data and part of single baryon 
properties and renders a mass of the H particle which agrees with the available
experimental data [17-19,23-25].

\noindent

\section{Results and discussions}

\noindent

We are now ready to study the structure of $\Delta\Delta$ dibaryons. First of
all, let us look at the symmetry character of the $\Delta\Delta$ system. As is
emphasized above, the antisymmetrization of quarks belonging to different 
clusters plays a substantial role in the structure of the dibaryon. In both
$\Delta\Delta_{(3,0)}$ and $\Delta\Delta_{(0,3)}$ systems, the expectation 
value of he antisymmetrizer in the spin-flavor-color space, $\langle {\cal A}
^{sfc}\rangle$, is equal to 2. As a result, the kinetic energy term provides 
an attractive effect. Therefore, although the CMI of $\Delta\Delta_{(0,3)}$ 
does not have an attractive nature like $\Delta\Delta_{(3,0)}$, it is still
possible to form bound $\Delta\Delta_{(0,3)}$ like $\Delta\Delta_{(3,0)}$.

\vspace{0.3cm}

With parameter Sets 1 and 2, we calculate the binding energy and the 
root-mean-square-radii (RMS) of two systems. The results are tabulated in 
Table 2.

\vspace{0.3cm}

It is shown that both $\Delta\Delta_{(3,0)}$ and $\Delta\Delta_{(0,3)}$ are
bound states, their binding energies are a few tens of MeV  and the
correspondent RMS' are around 1 fm. The binding nature 
of $\Delta\Delta_{(0,3)}$ is consistent with our speculation.
The binding energy of $\Delta\Delta_{(0,3)}$ is lower than that of 
$\Delta\Delta_{(3,0)}$ by 5-6 MeV. This is attributed
to the fact that the CMI of $\Delta\Delta_{(0,3)}$ is repulsive in contrast to that of
$\Delta\Delta_{(3,0)}$, although their antisymmetrization characters are
the same. The results with Sets 1 and 2 have small difference mainly because
of the variation of $g_q$ brought by the mass changes of $\kappa$ and 
$\epsilon$, although the strange meson clouds are not important in the 
non-strange system.

\vspace{0.3cm}
Then, we try to reveal the effects of chiral-quark interactions in
$\Delta\Delta$ systems. This can be realized by locating calculations in 
other two chiral quark models, the extended $SU(2)$ and $SU(2)$ chiral 
quark models. In the former one (hereafter call Model II), the 
chiral-quark interactions are provided by scalar meson $\sigma$ and all
pseudoscalar mesons $\pi$, $K$, $\eta$ and $\eta\prime$, while in the latter 
one (hereafter call Model III), the chiral-quark interactions are induced 
by the pseudoscalar meson $\pi$ and scalar meson $\sigma$ only.  
The results in these two models are also tabulated in Table 2. It is seen 
that the binding natures of the two systems remain the same, but the values of
binding energies and RMS' have substantial differences. In the higher spin system like $\Delta\Delta_{(3,0)}$, the binding energy would increase when
the model changes from I to II or III, and the largest value occurs in
Model II. But in the lower spin state like $\Delta\Delta_{(0,3)}$, the 
binding energy changes in the opposite direction and the largest value appears in Model I. Anyway, the fact that the binding
energy of the system with spin $S=3$ is larger than that of the system with 
spin $S=0$ remains un-changed when the model used is alternated. 

\centerline{\bf {Table 2~~~~Binding energy $B_{AB}$ and RMS for $\Delta\Delta_{(3,0)}$
 and $\Delta\Delta_{(0,3)}$ systems$^{\dag}$}}

\begin{center}
\begin{tabular}{|c|c|c|c|c|}
\hline
$~$          & \multicolumn{2}{|c|}{one channel}
& \multicolumn{2}{|c|}{one channel} \\
 Channel  & \multicolumn{2}{|c|}{$\Delta\Delta_{(3,0)}$}
 & \multicolumn{2}{|c|}{$\Delta\Delta_{(0,3)}$} \\
\cline{2-5}
  & $B_{\Delta\Delta_{(3,0)}}$ & $RMS$ & $B_{\Delta\Delta_{(0,3)}}$ & $RMS$ \\
  &$(MeV)$ & $(fm)$ & $(MeV)$ & $(fm)$   \\  \hline
Model I  Set1 &  22.2 & 1.01 &  16.0 & 1.10  \\  \hline
Model I Set2 &  18.5 & 1.05  &  13.5 & 1.14  \\  \hline
Model II &  64.8 & 0.84   & 6.3  & 1.25  \\ \hline
Model III    &  62.7 & 0.86   &  13.2 & 1.11  \\ \hline
\end{tabular}
\end{center}
$\dag$ {\footnotesize {$B_{AB}$ denotes the binding energy between
A and B baryons and RMS represents the corresponding root-mean-square
radius.}}

\vspace{0.3cm}

This result can be understood in the following way. The basic observation is that their symmetry structures, which cause $\langle {\cal A} \rangle^{sfc}=2$,
where $sfc$ denotes that the operation takes place in the spin-flavor-color
space only, is very beneficial in forming bound state. Namely, the quark
exchange effect due to the Pauli principle is enormous so that the quarks
in two clusters can be sufficiently close, and consequently the contribution from the kinetic energy term shows the attractive character. On the other hand,
the $\sigma$ field induced interaction provides a fairly strong attraction
which plays a dominant role in forming bound state. These two factors make 
the binding nature of the system stable with respect to models and
parameters.

\vspace{0.3cm}

Moreover, from the form of Gell-Mann matrices we know that
the K and $\kappa$ meson exchanges do not exist between u or d quark pairs,
namely the chiral fields K and $\kappa$ do not contribute in both systems.
The contributions from pseudoscalar mesons in two systems are comparable.
But the contribution from the $\sigma\prime$ meson are characteristically 
different in these two systems. Opposite to the contributions of CMI, it is
strongly repulsive in $\Delta\Delta_{(3,0)}$ and relatively weakly attractive
in $\Delta\Delta_{(0,3)}$, and the repulsive strength of $\sigma\prime$ is
much larger than the attractive strength in CMI. As a result, when one moves
from Model I to Model II or III, the scalar mesons $\sigma\prime$, $\kappa$ 
and $\epsilon$ are turned off, the wavefunction of $\Delta\Delta_{(3,0)}$
would move inward and that of $\Delta\Delta_{(0,3)}$ would distribute more
outward. Consequently, the contribution from $\sigma$ becomes stronger in 
$\Delta\Delta_{(3,0)}$ and a little weak in $\Delta\Delta_{(0,3)}$. That is
why when model alternates from I to II or III, $\Delta\Delta_{(3,0)}$ becomes
much more bounder and $\Delta\Delta_{(0,3)}$ turns out to be a little less 
bound. 

\vspace{0.3cm}

Above results can be crosschecked by their S-wave scattering phase shifts 
and corresponding scattering lengths. We plot the S-wave phase shifts 
of these two systems in Figs. 1 and 2, respectively. The solid and dashed
curves are those with parameter Sets 1 and 2, accordingly. With these 
phase shifts, one can easily extract the corresponding scattering lengths,
which are tabulated in Table 4.

\centerline{\bf {Table 4~~~~The Scattering length $a$ of the $\Delta\Delta$ systems}}
\begin{center}
\begin{tabular}{|c|c|c|}
\hline
        &   one channel          &     one channel\\
        &   $\Delta\Delta_{(3,0)}$  & $\Delta\Delta_{(0,3)}$\\ \hline
Model I  Set1 &   $-1.30~(fm)$         &     $-2.56~(fm)$\\ \hline
Model I   Set2 &   $-1.16~(fm)$         &     $-2.72~(fm)$\\ \hline
\end{tabular}
\end{center}
  Both phase shifts and scattering lengths are consistent with our
above results.

\noindent

\section{Conclusions}

In the $\Delta\Delta$ system, there exist two bound states,
$\Delta\Delta_{(3,0)}$ and $(d^*)$  $\Delta\Delta_{(0,3)}$. The former one 
has been predicted as a bound dibaryon [5,8-10]. Its binding energy varies
model by model. By employing the $SU(3)$ chiral quark model with which the
available empirical data can be well-reproduced, the binding energy of the 
one channel $\Delta\Delta_{(3,0)}$ (or $d^*$) is 18.5-22.2 MeV and the corresponding RMS is about 1.0 fm. This result is consistent with most
reports of other theoretical predictions [5,8-10]. Then, we predict the 
binding energy of $\Delta\Delta_{(0,3)}$. The result turns out to be
13.5-16.0 MeV and the corresponding RMS is around 1.1 fm. The binding 
behaviors of these systems are attributed
to their special symmetry properties which make the contribution of kinetic
energy term beneficial for forming dibaryons. Moreover,  
the chiral fields also provide substantial contributions to their binding
behaviors. Among these chiral fields, the relatively stronger $\sigma$ induced interaction is always the dominant one, no matter which model and model parameters are used. As a consequence, the binding phenomena of the  
$\Delta\Delta_{(3,0)}$ and $\Delta\Delta_{(0,3)}$ systems are always remained
the same. 

\vspace{0.3cm}

However, the mass of these two systems predicted in the one-channel 
approximation are all larger than the thresholds of strong
decay channels $\Delta\Delta\rightarrow{\Delta}N\pi$ and
$\Delta\Delta\rightarrow{NN}\pi\pi$ which are about 155 MeV and 310 MeV
( even after considering the $CC$ channel coupling in $d^*$, this conclusion 
is still true). These dibaryons would have very broad widths. Because the 
newly predicted $\Delta\Delta_{(0,3)}$ has the lower spin $S=0$ with 
respect to $\Delta\Delta_{(3,0)}$, it might be more
favorable to be experimentally detected.

\end{document}